\documentclass{article}
\usepackage{spconf, graphicx}
\usepackage{amsmath}
\usepackage{amssymb}
\usepackage{booktabs}
\usepackage{multirow}

\usepackage{enumitem}
\setlist{nosep, leftmargin=14pt}

\usepackage{mwe} 
\usepackage{caption}
\usepackage{subcaption}
\usepackage{array}
\usepackage{tikz}

\newcommand\copyrighttext{%
  \footnotesize \copyright  2023 IEEE.  Personal use of this material is permitted.  Permission from IEEE must be obtained for all other uses, in any current or future media, including reprinting/republishing this material for advertising or promotional purposes, creating new collective works, for resale or redistribution to servers or lists, or reuse of any copyrighted component of this work in other works.}
\newcommand\mycopyrightnotice{%
\begin{tikzpicture}[remember picture,overlay]
\node[anchor=south,yshift=10pt] at (current page.south) {\fbox{\parbox{\dimexpr\textwidth-\fboxsep-\fboxrule\relax}{\copyrighttext}}};
\end{tikzpicture}%
}

\title{Deep learning for predicting metastasis on melanoma WSIs}
\name{Christopher Andreassen$^1$*, Saul Fuster$^1$*, Helga Hardardottir$^{2,3}$, Emiel A.M. Janssen$^{2,3}$, Kjersti Engan$^1$\thanks{*These authors contributed equally.}}
\address{\small{$^1$ Dept. of Electrical Engineering and Computer Science, University of Stavanger, 4021 Stavanger, Norway} \\
\small{$^2$ Dept. of Pathology, Stavanger University Hospital, 4011 Stavanger, Norway} \\
\small{$^3$ Dept. of Chemistry, Bioscience and Environmental Engineering, University of Stavanger, 4021 Stavanger, Norway}}
\begin{document}
%
\maketitle
\mycopyrightnotice

\begin{abstract}
Northern Europe has the second highest mortality rate of melanoma globally. In 2020, the mortality rate of melanoma rose to 1.9 per 100 000 habitants. Melanoma prognosis is based on a pathologist's subjective visual analysis of the patient's tumor. This methodology is heavily time-consuming, and the prognosis variability among experts is notable, drastically jeopardizing its reproducibility. Thus, the need for faster and more reproducible methods arises. Machine learning has paved its way into digital pathology, but so far, most contributions are on localization, segmentation, and diagnostics, with little emphasis on prognostics. This paper presents a convolutional neural network (CNN) method based on VGG16 to predict melanoma prognosis as the presence of metastasis within five years. Patches are extracted from regions of interest from Whole Slide Images (WSIs) at different magnification levels used in model training and validation. Results infer that utilizing WSI patches at 20x magnification level has the best performance, with an F1 score of 0.7667 and an AUC of 0.81.
\end{abstract}



\section{Introduction}
\label{sec:intro}

The incidence rate of melanoma in Norway is among the highest in the world, steadily increasing yearly from 1961 to 2020. During the 2016-2020 quinquennial, the incidence rate increased by approximately 11\% for both sexes \cite{CancerRegistryNorway_2020}. Carefully analyzing cancer tumors for extracting clinically relevant diagnostic and prognostic information is important for treatment management and can be labor intensive for pathologists.  In general, the number of biopsies arriving at a hospital's pathologist's department requiring analysis by pathologists is increasing yearly \cite{weller2012aarhus}. Pathological prognosis of melanoma is based on the eight edition American Joint Committee on Cancer (AJCC) tumor, node and metastasis (TNM) staging system \cite{amin2017eighth}. The histopathological prognostic factors used in the system are subject to interobserver variability. Tumor thickness and ulceration are the most important prognostic factors \cite{balch2001prognostic}. A study on interobserver variability of histopathological factors found concordance between general pathologists and pathologists with expertise to be good overall, however it showed that variability of tumor thickness and ulceration resulted in reclassification of 15,5\% of thin melanomas \cite{eriksson2013interobserver}. This high level of uncertainty can only be confirmed by the development of the tumor over time. Combined with the increased incidence rate of melanoma, it is necessary to implement faster and more consistent methods for melanoma prognosis assessment.


\begin{figure*}[t!]
  \centering
  \includegraphics[width=0.95\linewidth, height=2in]{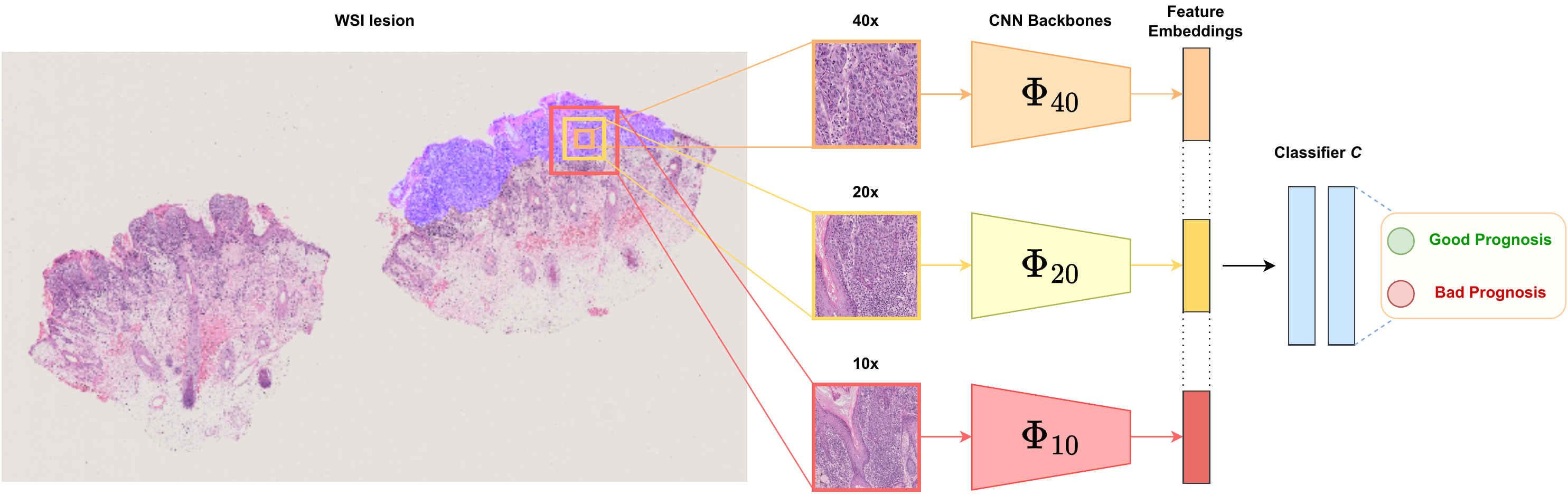}
\caption{Overview of a multiscale model for predicting melanoma prognosis. Patches from the defined lesion areas are extracted at different magnification levels and fed into independent CNN backbones $\Phi_{x}$. The extracted features from all CNNs are later concatenated and fed into a classifier $C$. For mono scale models, a single backbone is used and the output feature embedding is directly fed into $C$.}
\label{fig:OverwiewPipeline}
\end{figure*}

Computational pathology (CPATH) systems can be used for fast and reproducible analysis and reduce pathologists workload for example as decision support tools \cite{cui2021artificial}. Machine learning (ML) has grown in popularity due to the increase of available data and processing capabilities. Most published works in CPATH is in the field of diagnostics, detecting tumor areas \cite{kather2016multi}, grading tumor \cite{wetteland2021automatic}, among others. 
Bhattacharjee et al. \cite{bhattacharjee2020efficient} proposed a method using two CNNs and ensemble learning for classifying prostate tumors as benign or malignant.
Wang et al. \cite{wang2020breast} proposed a method that classified regions in breast tumors 
utilizing four pre-trained deep neural networks to increase the robustness of the method.  
In a similar fashion, multi-scale models have also been successfully used for computer-aided diagnosis systems in order to combine global and local patterns extracted at different magnification levels \cite{wetteland2021automatic,fuster2022invasive}.

CPATH for prognostics is more challenging as it attempts to detect future events. Although certain factors lead to worse outcomes, the relationship between such factors and the patient's outcomes is not causal \cite{scolyer2020melanoma}. Moreover, there is not necessarily a region in the WSI that may be an indicator for providing a reliable prognosis.  However, there are some works in the literature estimating prognostic values from WSI. Prognosis identification has been done in screening programs, breast cancer screening and cervical screening \cite{cruz2006applications, kourou2015machine}. Dlamini et. al. \cite{dlamini2020artificial} claims that utilizing ML to analyze WSIs can help pathologists find a likely prognosis of malignant tumors, leading to reproducible and faster evaluation of the tumors. Although ML methods for predicting prognosis of melanoma using dermoscopic images does exist \cite{ winkler2020melanoma}, no prognostic methods for melanoma that utilizes histological WSIs were found in the existing literature.

In this work, we present a convolutional neural network (CNN) method based on VGG16 for predicting the prognosis of melanoma from histopathological WSIs. This method exploits a multi-scale multi-input CNN backbone that aggregates the image features extracted from patches at different magnification levels. We compare the performance of models that combine from one up to three magnification levels, along with different patch extraction configurations. We test our models with a private cohort of melanoma WSIs. 



\section{Data Material}
\label{sec:materialsandmethods}
\noindent
The data material in this work consists of 52 WSIs from 52 patients, all with verified malignant melanoma and 5 years follow-up, diagnosed at Stavanger University Hospital between 2008-2012. The data is balanced in terms of outcome, as 26 of the patients are considered to have a good prognosis and the remaining 26 to have a bad prognosis. The prognostic outcome was determined based on the presence of a local or distant metastasis or no metastasis within five years. The WSIs were stained with Haematoxylin and Eosin (H\&E) stain and scanned at 40x magnification with a Hamamatsu Nanozoomer s60 scanner, and stored in NDPI format. 

\begin{table}[htb]
\begin{center}
\small
\scalebox{0.83}{
\begin{tabular}{|p{1cm}|p{1.3cm}|l|}
    \hline
    Variable & Values & Description \\\hline\hline
    $m$ & $10$, $20$, $40$ & Magnification level(s) of extracted start coordinates. 
    \\\hline
    $t/v$ & $t$, $v$ & $t$ indicates training dataset and $v$  validation dataset. \\\hline
    $k$ & $k\in \mathbb{N}$ & Iteration nr. during  crossvalidation. \\\hline
\end{tabular}}
\caption{Variables used to describe a dataset $D^{m}_{t/vk}$.}
\label{tab:listVariables}
\end{center}
\end{table}

In all 52 WSIs, the lesion was annotated manually by a pathologist (HH). The annotation protocol was to annotate the lesions on one of the sections in all patients. Some areas of normal epithelial tissue and other typical structures were also annotated, but not necessarily in all images. All annotations were done roughly, meaning they are not very detailed on the edges. 
Rough annotations provide annotations on the entire dataset within a reasonable time and work effort.

\section{Methods }
\subsection{Notation} 

\noindent
Let $I_{WSI}^{x}$ correspond to a WSI at magnification level $x$.  $I_{\text{WSI}}^{40}$ are very large gigapixel images, and it is not feasible to process the entire WSI at once. As such, all  CPATH systems resort to patching or tiling of the image, or the region of interest in the image, before further processing. Let
\begin{equation}
    {\cal T} : I_{\text{WSI}\in R}^{x} \rightarrow \{I_{p}^{x}; p=1\cdots \}
    \label{eq:region}
\end{equation} represent the process of tiling a region defined by $R$ of the image $I_{\text{WSI}}^{x}$ into a set of patches.

A parameterized patch extraction algorithm proposed by Wetteland et.al. \cite{wetteland2021parameterized} is used in this paper, conveying how the algorithm extracts coordinates from one magnification, representing patches in a WSI. First, 20x magnification is used to define valid patches inside the region $R$. Then, the center pixel of the patch is projected to other magnifications to maintain the same physical midpoint for all views. This way, we ensure that the view is centered regardless of the magnification level of choice. The patch size remains the same for all magnifications. Therefore, decreasing the magnification level will widen the field of view, as illustrated in Figure \ref{fig:OverwiewPipeline}.
A variable $D$ refers to a dataset in the form of start coordinates. The datasets are defined by variables shown in Table \ref{tab:listVariables}, with the format $D^{m}_{t/vk}$.

\subsection{Preprocessing}

\begin{figure}[hbt!]
  \centering
  \includegraphics[width=\linewidth, height=1.6in, trim={0 7.7cm 0 0},clip]{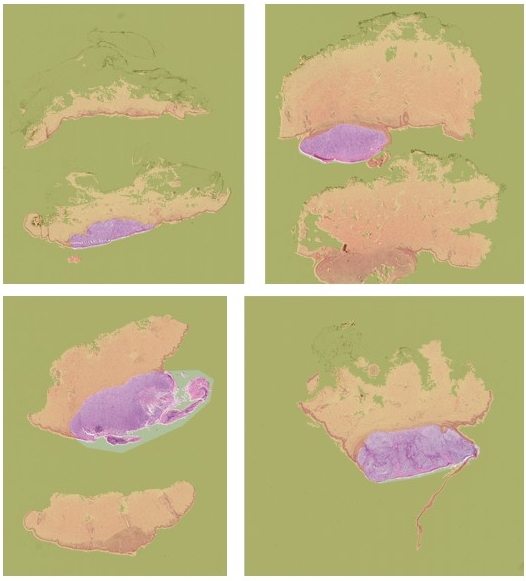}
  \caption{Examples of tissue masks and annotated masks showing lesions in the WSIs. The area shaded with green is outside the tissue mask, and the area shaded with orange is outside the annotated mask. The area inside both masks contains the lesion.}
  \label{fig:TissueMasks_withTissue}
\end{figure}

\noindent
Several preprocessing steps were applied to all 52 WSIs. Region of interest masks were extracted from areas annotated by the pathologist. Then, tissue masks were generated using HSV format to locate the blue and magenta color range, corresponding to H\&E stained tissue areas. The color range, hue, of WSIs in the HSV format was set to 100-180. Overlapping areas of tissue and region of interest masks resulted in a lesion mask, a segmented lesion area, without the background noise located outside of the tissue mask. Closing and opening morphological operations were applied to lesion masks to remove small regions and fill small holes, respectively, giving the final region $R$ used in further processing, see Eq\ref{eq:region}. Some examples of segmented lesions are displayed in Figure \ref{fig:TissueMasks_withTissue}.

\subsection{Patch based prediction}
\noindent
The proposed method for patch based prediction is illustrated in Figure \ref{fig:OverwiewPipeline}. 
Pathologists examine malignant tissue at multiple levels to assess the prognosis of a patient. Mimicking this, multiple magnification levels were used to predict prognosis of melanoma. This was done using CNN backbone for mono-scale (MONO) single input using patches extracted at particular magnification level; or multi-scale multi-input CNN backbone, that would range from di (DI) to tri-scale (TRI), 
 inspired by Wetteland et.al. \cite{wetteland2021automatic}.

Let $ f_{p}^{x}=\Phi_{x}(I_{p}^{x}) \subset \mathbb{R}^{512}$
represent the feature embedding of patch $p$ at magnification level $x \in \{10,20,40\}$. 
At monoscale, the feature embedding is further passed through a patch-based classifier, $C$, giving a binary prediction:  $y_{p}^{x} = C_{x} (f_{p}^{x}) = C_{x}(\Phi_{x} (I_{p}^{x}))$, for $y_{p}=1$ corresponds to bad prognosis, and vice versa. For the multiscale approach, a total feature vector is found by concatenating the feature embeddings from the used magnification levels per patch: 
$\bar{f}_{p}=[f_{p}^{x_{1}T}f_{p}^{x_{2}T} \cdots]^{T}$ resulting in a feature vector of size $m \times 512$ where $m$ is the number of scales. The classifier(s) $C_{x}$ and 
$C_{x_{1},x_{2} \cdots}$ are all fully connected networks consisting of two dense layers of 4096 neurons each and a third dense layer as a binary output 
$y_{p}^{x1,x2 \cdots} = C_{x1,x2, \cdots} (\bar{f}_{pt})$. The size of input layers varies with the number of scales, while the output layer remains in two neurons for the good and bad prognosis prediction. Patch based prediction is straight forward, but problematic in the sense that no real truth data exist. We know truth data at a patient level, and for patch based prediction, we let all patches inherit the truth label of the patient. 

\subsection{Patient based prediction}
We propose to find a patient-prediction $Y$ from all the patch predictions $y_{p}$ within the $R$ region from a WSI.  We do not know if scattered information in the lesion is a stronger or weaker indication of bad prognosis compared to localized information, thus we propose a simple model counting the total number of bad and good patch-predictions for patient P, $Y=\frac{1}{N_R}\sum_{p \in R}y_{p}^{x}$, and $ Y>T \rightarrow \text{bad prognosis}$, where $N_R$ is the number of patches within a region $R$, and $T$ is a threshold that we will estimate from a receiver operating characteristic (ROC) curve.



\section{Experiments}
\label{sec:experimental setup}

Experiments were done for mono scale for the patch-based prediction, and both mono and multi-scale for patient based prediction. Training and validation datasets were created on a patient basis, using stratified 5-fold cross validation, using up to 250 patches per WSI. 
The feature extraction layers of the network $\Phi_{x}$ were frozen, leaving only the classifier with trainable parameters. The CNN backbones used in this work are pre-trained VGG16 networks from ImageNet \cite{krizhevsky2017imagenet}. The number of trainable parameters of the classifiers $C_{x}$ were kept low to prevent overfitting. The classifiers were trained up to 20 epochs, until the validation loss converged. A stochastic gradient descent optimizer was used during training with 0.9 momentum, while learning rate varied for mono and multi-scale models. Random resize, crop and random horizontal flip were used for augmentation in the training set, while the dropout rate was set to 50\%. Early stopping was used if the validation loss did not converge over time.

\begin{figure*}[htb]
    \centering
    \begin{subfigure}[b]{0.33\textwidth}
        \centering
        \includegraphics[width=\textwidth, height=1.4in]{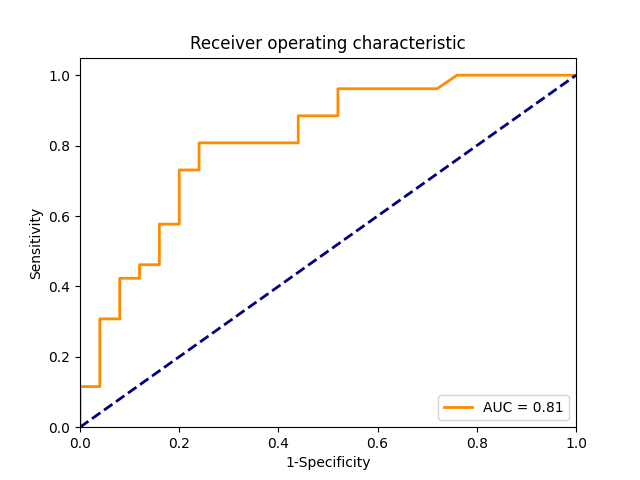}
        \caption{MONO$_{20x}$.}
        \label{fig:ROC_20x_OT20x}
    \end{subfigure}
    \hfill
    \begin{subfigure}[b]{0.33\textwidth}
        \centering
        \includegraphics[width=\textwidth, height=1.4in]{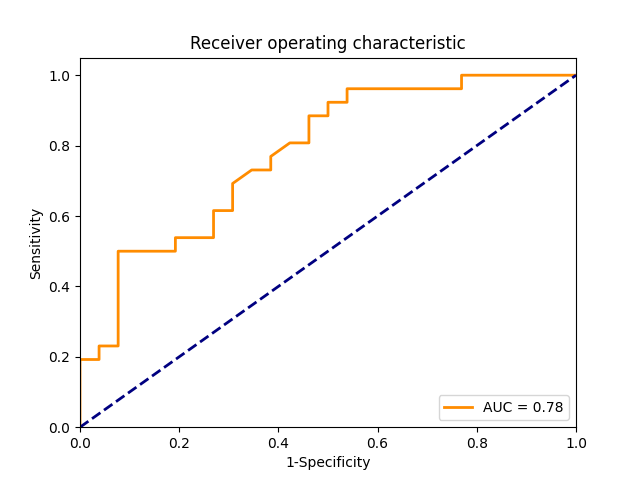}
        \caption{DI$_{20,40x}$ .}
        \label{fig:ROC_20x40x_OT20x}
    \end{subfigure}
    \hfill
    \begin{subfigure}[b]{0.33\textwidth}
        \centering
        \includegraphics[width=\textwidth, height=1.4in]{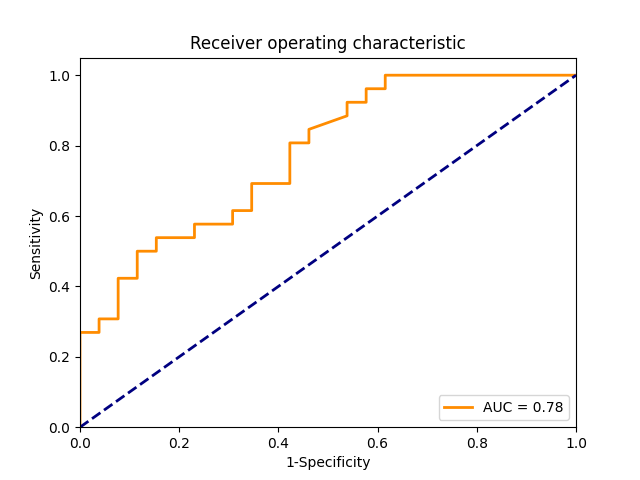}
        \caption{TRI$_{10,20,40x}$.}
        \label{fig:ROC_10x20x40x_OT20x}
    \end{subfigure}
    \caption{ROC curve and AUC from patient-based prediction. From left to right, the plots correspond to MONO-$20$, DI$_{20,40x}$ and TRI model architectures.}
    \label{fig:ROC}
    
\end{figure*}




For patient based prediction, the threshold $T$ was found from ROC curves based on the best trade-off between highest sensitivity and specificity on the training set, tested on the validation set using cross-validation. This experiment was done with a learning rate of 0.0001 for MONO and 0.001 for DI and TRI. Cross-validation was done using patches from magnification levels ($m$) $20$, $20-40$ and $10-20-40$, for MONO, DI and TRI model architectures respectively. We have not found any relevant work in the literature to compare with, predicting prognostics from WSI with melanoma.


\section{Results \& Discussion}
\label{sec:resultsanddiscussion}

Results from patch-based experiments are shown in Table \ref{tab:Results20x_Mono}. Patch-based prognosis prediction shows promising results for models MONO$_{20x}$ and MONO$_{40x}$. Models can differentiate between patches from WSIs with good and bad prognoses. We observe that models tend to present higher sensitivity over specificity. MONO$_{20x}$ has the highest F$_1$ score and relatively high sensitivity and specificity. This may due to (m) $20$ reaching a compromise between gathering enough contextual information from neighbouring tissue and maintaining a certain level of detail from local cellular patterns. Another reason for high sensitivity rates is the distribution of genuine bad prognosis patches. While WSIs labeled with a good prognosis are populated by tissue presenting characteristic features of positive outcomes, the same does not hold true for bad prognosis. In bad prognosis WSIs, there might be patches that present positive prognostic features as well as negatives, and as a result, the number of actual bad prognosis patches is generally underrepresented. Moreover, which of those patches present actual bad prognosis features is unknown, which is reflected in the obtained metrics.

\begin{table}[htb]
\begin{center}
\small
\scalebox{0.85}{
\begin{tabular}{|c|c|c|c|c|c|}
        \hline
            Architecture & Validation dataset & Sensitivity & Specificity & F$_1$ score \\\hline\hline
             \multirow{3}{*}{MONO} & $D^{10}_{v1}$ & 0.6916 & 0.5576 & 0.6699 \\\cline{2-5}
              & $D^{20}_{v1}$ & 0.7928 & \textbf{0.5824} & \textbf{0.7392} \\\cline{2-5}
              & $D^{40}_{v1}$ & \textbf{0.8205} & 0.4552 & 0.7197 \\
        \hline
        \end{tabular}}
        \caption{Results from patch-based prediction, using MONO model (one fold).}
        \label{tab:Results20x_Mono}
\end{center}
\end{table}

Results from patient-based experiments are shown in Table \ref{tab:combinedpatient}, for MONO, DI and TRI models respectively. In the likes of the patch-based experiment, we can observe that models do also tend to have higher sensitivity over specificity. This pattern holds true for all iterations, regardless of the model's architecture. DI$_{20,40x}$ are the least sensitive, followed by MONO$_{20x}$ and TRI$_{10,20,40x}$  in descending order. The aforementioned class imbalance is also reflected in the threshold, which requires a minority of patches to predict a global bad prognosis label. Plots showing the ROC curve and AUC for this experiments are plotted in Figure \ref{fig:ROC}, where MONO$_{20x}$ obtained the highest AUC. Patient-based prognosis prediction performs highly for all model architectures. Our best performing model MONO$_{20x}$ shows an AUC, F1 score, and accuracy of 0.81, 0.7667, and 0.7255, respectively.

\begin{table}[htb]
\begin{center}
\small
\scalebox{0.82}{
\begin{tabular}{|c|c|c|c|c|c|}
        \hline
            Model & Threshold & Sensitivity & Specificity & F$_1$ score & Accuracy \\\hline\hline
            MONO$_{20x}$ & 0.3720 & \textbf{0.8846} & \textbf{0.5600} & \textbf{0.7667} & \textbf{0.7255} \\
            DI$_{20,40x}$ & 0.4720 & \textbf{0.8846} & 0.5385 & 0.7541 & 0.7115 \\
            TRI$_{10,20,40x}$  & 0.3240 & 0.8462 & 0.5385 & 0.7333 & 0.6923 \\
        \hline
        \end{tabular}}
        \caption{Results from patient-based prediction. Metrics reflect the mean values of models trained using cross validation.}
        \label{tab:combinedpatient}
\end{center}
\end{table}

Although pathologists can look upon clinical data and the entire tissue block for predicting melanoma prognostics, MONO$_{20x}$ has shown good performance using a single WSI per patient. AUC of 0.81 is promising when taken into account the reported interobserver variability among expert pathologists for evaluating prognostic parameters\cite{eriksson2013interobserver}. The reason for multi-scale models to perform worse might be more trainable parameters, and a small training set.


\section{Conclusion}
\label{sec:conclusion}

In this paper, we show that CNNs can be used to predict the prognosis of melanoma as a proof of concept with a small dataset. A multi-scale multi-input backbone was implemented to leverage information conveyed at different magnification levels, but we found that mono-scale 20x magnification gave the most promising results with F1 score of 0.7667 and an AUC of 0.81. 20x seems to provide a good trade-off between local and global information. In future work larger datasets should be tested both to explore if that can make multi-scale models perform better, and to verify the encouraging overall prognostic prediction results.

\section{Compliance with ethical standards}
\label{sec:compliance}

This study was performed in line with the principles of the Declaration of Helsinki. Approval was granted by the Regional Ethics Committee (No: 2019/747/RekVest). The authors have no relevant financial or non-financial interests to disclose.

\section{Acknowledgements}
\label{sec:acknowledgements}

This research has received funding from the European Union's Horizon 2020 research and innovation program under grant agreements 860627 (CLARIFY) and “Pathology services in the Western Norway Health Region – a centre for applied digitization” from a Strategic investment from the Western Norway Health Authority.



\bibliographystyle{IEEEbib}
\bibliography{strings,refs}

\end{document}